\documentclass[aps,pra,twocolumn,superscriptaddress,amsmath,amssymb,showpacs,notitlepage,longbibliography]{revtex4-1}
\pdfoutput=1
\usepackage{algorithm}
\usepackage{color}
\usepackage{algorithmic}
\usepackage{amsmath}
\usepackage{braket}
\usepackage{multirow}
\usepackage{amssymb}
\usepackage{amsthm}
\usepackage{bm}
\usepackage{thmtools}

\usepackage{array}
\usepackage{url}
\usepackage{hyperref}
\hypersetup{
    colorlinks=true,
    linkcolor=blue,
    filecolor=magenta,      
    urlcolor=cyan,
    citecolor=red
}
\usepackage{graphicx}
\usepackage{bibentry}
\usepackage{booktabs}
\usepackage{dcolumn}
\usepackage{bm}
\usepackage{blkarray}
\usepackage{xparse}
\usepackage{mathtools}
\usepackage{microtype}
\usepackage{enumerate}
\usepackage{soul}

\usepackage{makecell}

\providecommand{\U}[1]{\protect\rule{.1in}{.1in}}
\newtheorem*{theorem*}{Theorem}

\newcommand{\CG}[1]{{\color{red}#1}}

\begin{document}

\title{Bayesian quantum estimation of the separation of two incoherent point sources}

\author{Boyu Zhou}
\affiliation{Department of Physics, The University of Arizona, Tucson, Arizona, 85721, USA}

\author{Saikat Guha}
\affiliation{Department of Electrical and Computer Engineering, University of Maryland, College Park, Maryland, 20742, USA}
\affiliation{Wyant College of Optical Sciences, The University of Arizona, Tucson, Arizona, 85721, USA}

\author{Christos N. Gagatsos}
\affiliation{Department of Electrical and Computer Engineering, The University of Arizona, Tucson, Arizona, 85721, USA}
\affiliation{Wyant College of Optical Sciences, The University of Arizona, Tucson, Arizona, 85721, USA}
\affiliation{Program in Applied Mathematics, The University of Arizona, Tucson, Arizona 85721, USA}

\begin{abstract}
We address the estimation problem of the separation of two arbitrarily close incoherent point sources from the quantum Bayesian point of view, i.e., when a prior probability distribution function (PDF) on the separation is available. 
For the non-dispalced and displaced half-Gaussian prior PDF, we compare the performance of SPADE and direct imaging (DI) with the Bayesian minimum mean square error and by varying the prior PDF's parameters we discuss the regimes of superiority of either SPADE or DI.
\end{abstract}
\maketitle

\section{Introduction}

In recent years, the rapid development of quantum information theory provided new perspectives and enhanced the performance of information science tasks. An example of paramount importance is the Rayleigh limit or diffraction limit~\cite{born2013principles,rayleigh1879xxxi} which indicates that two point sources cannot be resolved by direct imaging (DI) when the separation is below the diffraction-limited spot size of the point-spread function (PSF) of the imaging system. This imposed a limitation on astronomical observation technology and microscopy. However, based on quantum and classical Fisherian estimation theory, Tsang \textit{et al}~\cite{PhysRevX.6.031033} showed that the Rayleigh limit can be exceeded by using a spatial-mode demultiplexing (SPADE) measurement. Further research on said topic includes generalized to arbitrary (non-Gaussian) PSFs \cite{kerviche2017fundamental, tsang2018subdiffraction}, estimation of multi points or extended objects \cite{Bisketzi_2019, dutton2019attaining, prasad2020quantum, sajjad2024quantum}, estimation of multiple parameters \cite{vrehavcek2017multiparameter, prasad2019quantum, Prasad_2020}, estimation in the two and three dimensions \cite{ang2017quantum, yu2018quantum, napoli2019towards}, and adaptive methods \cite{Bao:21, matlin2022imaging, lee2022quantum}.

The literature so far is populated with works based on Fisherian approach, i.e., when the estimated parameter has a fixed, yet unknown, value. This way, one can establish a lower bound on the the covariance matrix of any unbiased estimator by computing the quantum and classical Cram\'er-Rao bounds (CRB), given by the inverse of the quantum Fisher information and its classical counterpart, the classical Fisher information. 

The Fisherian approach requires a large number of independent sensing attempts so that the Fisherian CRB is asymptotically attained \cite{Morelli_2021}. If, instead of collecting data, we are presented with a prior probability distribution (PDF) on the unknown parameter, one can follow the Bayesian approach to evaluate the minimum mean square error (MMSE)  \cite{Personick1971} (instead of lower bounding the MSE). Said Bayesian MMSE is guaranteed to be always attainable by a single-shot projective measurement \cite{Personick1971}, while optimality of the MMSE is not improved if one considers positive operator-valued measures \cite[Appendix A therein]{Macieszczak2014}. In this work we follow the Bayesian approach for estimating the separation of two incoherent point sources with a Gaussian PSF, assuming that the prior PDF is the half-Gaussian.

This paper is organized as follows: In Sec.~\ref{SEC: Quantum description}, we describe the physical setting. In Sec.~\ref{SEC: Bayesian quantum parameter estimation} we briefly give the mathematical tools we utilize for computing the MMSE. For the non-displaced half-Gaussian prior PDF, in Sec.~\ref{SEC: MMSE of two incoherent optical point sources} we compute the MMSE for the separation of two incoherent point sources and in Sec.~\ref{SEC: Measurement technology} we compute the MSEs for DI and SPADE. In Sec.~\ref{SEC: Displaced half Gaussian PDF} we compute and compare the MMSE, DI, and SPADE for the displaced half-Gaussian. Lastly, in Sec.~\ref{SEC: Conclusion} we discuss our results and further research directions.

\section{Quantum description of the source and imaging model}\label{SEC: Quantum description}
The quantum state of the two incoherent point sources is \cite{PhysRevLett.109.070503, PhysRevX.6.031033},
\begin{equation}
\hat{\rho}_{\text{opt}}=(1-\varepsilon)\hat{\rho}_{\text{vac}}+\varepsilon \hat{\rho} + \boldsymbol{O}(\varepsilon ^2),
\end{equation}
where $\hat{\rho}_{\text{vac}}=|0 \rangle \langle 0 |$ represents the vacuum state, $\hat{\rho}$ represents a single-photon state, while higher photon number contributions are negligible. Since the vacuum state $\hat{\rho}_{\text{vac}}$ does not carry information, we will be concerned with the state $\hat{\rho}$.

We make the following assumtions on the properties of the two-point source \cite{PhysRevX.6.031033, goodman2015statistical, mandel1995optical, labeyrie2006introduction, tsang2011quantum}: They are incoherent and located in a one-dimensional spatial configuration, they are weak enough so that the average photon number $\epsilon$ arriving on the image plane at a coherence time is $\epsilon << 1$. Also, the far-field radiation from sources which is collected at the entrance pupil of an optical imaging system (e.g. microscope or telescope) is assumed quasi-monochromatic, and the paraxial approximation is valid.

On the image plane, the one-photon state $\hat{\rho}$ related to the mutual coherence of the optical fields with respect to the Sudarshan-Glauber distribution, becomes,
\begin{equation}
\hat{\rho}= w_1 |\psi_1\rangle \langle \psi_1|+w_2|\psi_2\rangle \langle \psi_2|,
\end{equation}
where $w_1+w_2=1$, $w_i > 0$ is the relative intensity or brightness. The states $|\psi_i\rangle$ can be written as,
\begin{equation}
|\psi_i\rangle=\int dx \Psi_{\text{PSF}}(x-\chi_i)|x\rangle,
\end{equation}
where $\chi_i$ is the position of the source on the object plane and $\Psi_{\text{PSF}}(x)$ is the point-spread function (PSF). As pointed out in \cite{Bisketzi_2019}, an ideal imaging system with $\Psi_{\text{PSF}}(x)=\delta (x)$ is free of any Rayleigh limit. However, in practice and per our assumptions on quasi-monochromatic paraxial light, we will work with the Gaussian PSF,
\begin{equation}
\Psi_{\text{PSF}}(x)=\frac{1}{(2\pi \tilde{\sigma}^2)^{\frac{1}{4}}}e^{-\frac{x^2}{4\tilde{\sigma}^2}},
\end{equation}
where $\tilde{\sigma}=\lambda/(2 \pi \text{NA})$, $\lambda$ is the free space wavelength and $\text{NA}$ is the effective numerical aperture of the imaging system. The Gaussian PSF in conjuction with the Hermite-Gauss (HG) modes presents a mathematical convenience because the state $\hat{\rho}$ can be mathematically (yet not physically) represented as a statistical mixture of coherent states Refs.~\cite{Bisketzi_2019, dutton2019attaining, sajjad2024quantum}. Utilizing the HG modes, we write,
\begin{equation}
|\psi_i \rangle =\sum_{q=0}^{\infty}\langle \phi_q|\psi_i\rangle |\phi_q\rangle,
\end{equation}
where $|\phi_q\rangle$ is the HG mode which can be expressed in the position basis $|x\rangle$ as,
\begin{equation}
|\phi_q\rangle =\frac{1}{(2\pi \tilde{\sigma}^2)^{\frac{1}{4}} \sqrt{2^q q!}} \int dx H_q\left(\frac{x}{\sqrt{2\tilde{\sigma}^2}}\right) e^{-\frac{x^2}{4\tilde{\sigma}^2}} |x\rangle,
\end{equation}
where $H_q(\cdot)$ is the Hermite polynomial. Then, the coefficients can be written as,
\begin{eqnarray}
\nonumber \langle \phi_q|\psi_i\rangle &=&
\frac{1}{\sqrt{2\pi \tilde{\sigma}^2} \sqrt{2^q q!}} \int dx dx^{\prime} H_q\left(\frac{x}{\sqrt{2\tilde{\sigma}^2}}\right) e^{-\frac{x^2}{4\tilde{\sigma}^2}}\\
\nonumber &\times& e^{-\frac{(x^{\prime}-\chi_i)^2}{4\tilde{\sigma}^2}} \langle x^{\prime} |x\rangle\\
\nonumber &=&\frac{e^{-\frac{\chi_i^2}{8\tilde{\sigma}^2}}}{\sqrt{2\pi \tilde{\sigma}^2 2^q q!} }\int dx H_q\left(\frac{x}{\sqrt{2\tilde{\sigma}^2}}\right) e^{-\frac{(x-\chi_i)^2}{2\tilde{\sigma}^2}}\\
\label{eq:coefFock}&=&e^{-\frac{1}{2}(\frac{\chi_i}{2\tilde{\sigma}})^2} \frac{(\frac{\chi_i}{2\tilde{\sigma}})^q}{\sqrt{q!}}.
\end{eqnarray}
By setting $\frac{\chi_i}{2\tilde{\sigma}}=\alpha_i$, the coefficients of Eq. \eqref{eq:coefFock} take the form of projecting a coherent state on the Fock basis. Therefore, we can write the single-photon state (on the image plane) in the HG basis as,
\begin{equation}
\label{Eq: density operator in HG basis}
\hat{\rho} \equiv \hat{\rho}_{\alpha}= w_1 |\alpha_1 \rangle \langle \alpha_1|+w_2 |\alpha_2 \rangle \langle \alpha_2|,
\end{equation}
where $\alpha_i=\frac{\chi_i}{2\tilde{\sigma}} \in \mathbb{R}$. Therefore, our problem can be formulated as estimating the separation of coherent states.

\section{Bayesian quantum parameter estimation}\label{SEC: Bayesian quantum parameter estimation}
We assume that the two incoherent sources are centered symmetrically on either side of an origin, i.e., the centroid is assumed known and equal to zero. We also assume that the two point sources have equal brightness. Under these assumptions the density operator of Eq. \eqref{Eq: density operator in HG basis} becomes,
\begin{equation}
\label{Eq: centroid and equal brightness density operator in HG basis}
\hat{\rho} \equiv \hat{\rho}_{\alpha}= \frac{1}{2} |\alpha \rangle \langle \alpha|+\frac{1}{2} |-\alpha \rangle \langle -\alpha|,
\end{equation}
with $\alpha=\frac{\chi}{2\tilde{\sigma}} \geq 0 \in \mathbb{R}$. We note that we set $\alpha \geq 0$ so that each one of the two point sources have fixed position with the respect to the origin. Also, we note that the separation is in fact $2 \chi$. However, we will use the equivalent (real and non-negative) parameter $\alpha$ to simplify the calculation.

The Bayesian MSE is defined as \cite{Personick1971}:
\begin{eqnarray}
  \label{eq:deltaB}  \delta_B = \int_0^{\infty} d\alpha P(\alpha) \text{tr}\left[\hat{\rho}(\hat{H}-\alpha\hat{I})^2\right],
\end{eqnarray}
where $P(\alpha)$ is the prior PDF on the unknown parameter $\alpha$, $\hat{\rho}$ is the parameter-carrying state and $\hat{H}$ is the Hermitian operators whose eigenvectors provide the projective measurement. Then, the minimization is performed over all Hermitian operators $\hat{H}$, i.e.,
\begin{eqnarray}
\delta \equiv \underset{\hat{H}}{\text{min}}\ \delta_B,
\end{eqnarray}
where $\delta$ is the MMSE given by \cite{Personick1971},
\begin{eqnarray}
\label{eq:delta1}\delta = \text{tr}\hat{\Gamma}_2 - \text{tr}(\hat{B} \hat{\Gamma}_1),
\end{eqnarray}
where,
\begin{eqnarray}
  \label{eq:Gammak}  \hat{\Gamma}_k = \int_0^{\infty} d\alpha P(\alpha) \alpha^k \hat{\rho}(\alpha),\ k=0,1,2,
\end{eqnarray}
while the optimal $\hat{H}$ is denoted as $\hat{B}$ and is given by,
\begin{eqnarray}
\label{eq:B}\hat{B} = 2 \int_0^\infty dze^{-z \hat{\Gamma}_0} \hat{\Gamma}_1 e^{-z \hat{\Gamma}_0}
\end{eqnarray}
and the projective measurement consisting of the eigenvectors of $\hat{B}$ attain the MMSE of Eq. \eqref{eq:delta1}.

\section{MMSE of two incoherent point sources: Half-Gaussian prior PDF}\label{SEC: MMSE of two incoherent optical point sources}
For a coherent state $|\alpha \rangle$, the amplitude is $\alpha=\frac{1}{\sqrt{2}}(q_{\alpha}+ip_{\alpha})$, where we chose to work with $\hbar =1$. In our case $\alpha \geq 0$, therefore we set $p_{\alpha}=0$. For convenience, we change again the estimation parameter to $q_{\alpha}=\sqrt{2}\alpha$. 

Now, the task is to estimate the real and non-negative parameter $q_\alpha$, encoded in $\hat{\rho}$, when the prior probability distribution $P(q_\alpha)$ is the half-Gaussian distribution given by,

\begin{eqnarray}
\label{eq:HalfGaussian}	P(q_\alpha) = \frac{2}{\sigma \sqrt{2\pi}} e^{-\frac{q_\alpha^2}{2\sigma^2}},\ q_\alpha \geq 0,
\end{eqnarray}
where $\sigma$ is known. We note that the half-Gaussian PDF has support on the non-negatives and therefore the signs of the coherent states in Eq. \eqref{Eq: centroid and equal brightness density operator in HG basis} remain unchanged.

For this case Eq. \eqref{eq:Gammak} gives,
\begin{eqnarray}
\nonumber \hat{\Gamma}_k &=& \frac{2}{\sigma \sqrt{2\pi}}\int_{0}^{+\infty} dq_\alpha q_\alpha^k\\
\label{eq:GammakTP} &\times& e^{-\frac{q_\alpha^2}{2\sigma^2}} \left(\frac{1}{2}|\alpha\rangle\langle \alpha |+\frac{1}{2}|-\alpha\rangle\langle -\alpha |\right).
\end{eqnarray}
With a change of variables $q_\alpha \rightarrow -q_\alpha$, Eq. \eqref{eq:GammakTP} can be rewritten as,
\begin{eqnarray}
\nonumber \hat{\Gamma}_k &=& \frac{1}{\sigma \sqrt{2\pi}}\int_{0}^{+\infty} dq_\alpha q_\alpha^k e^{-\frac{q_\alpha^2}{2\sigma^2}}|\alpha\rangle \langle \alpha|\\
\label{eq:GammakTP2} &+&\frac{(-1)^k}{\sigma \sqrt{2\pi}} \int_{-\infty}^{0} dq_\alpha q_\alpha^k e^{-\frac{q_\alpha^2}{2\sigma^2}}|\alpha\rangle \langle \alpha|,
\end{eqnarray}
where we always set $\alpha=q_\alpha/\sqrt{2}\geq 0$.

For $k=0$ and $k=2$ Eq. \eqref{eq:GammakTP2} gives,
\begin{eqnarray}
\label{eq:Gamma0TP} \hat{\Gamma}_0 &=& \frac{1}{\sigma \sqrt{2\pi}}  \int_{-\infty}^{+\infty} dq_\alpha e^{-\frac{q_\alpha^2}{2\sigma^2}}|\alpha\rangle \langle \alpha|,\\
\label{eq:Gamma2TP} \hat{\Gamma}_2 &=& \frac{1}{\sigma \sqrt{2\pi}} \int_{-\infty}^{+\infty} dq_\alpha q_\alpha^2 e^{-\frac{q_\alpha^2}{2\sigma^2}}|\alpha\rangle \langle \alpha|,
\end{eqnarray}
respectively.

Here we note that Eqs. \eqref{eq:Gamma0TP} and \eqref{eq:Gamma2TP} have been considered in a different context in \cite{Personick1971} and \cite[Ch. VIII, Sec. 3 therein]{helstrom1976Book}. We also note that Eqs. \eqref{eq:Gamma0TP} and \eqref{eq:Gamma2TP} would be the same as if we were working on localizing a single point source. In any case, we present our own calculation in Appendix~\ref{App: Single source} in which we find, 
\begin{eqnarray}
\label{eq:Gamma0Amp2} \hat{\Gamma}_0= (1-s)\sum_{n=0}^\infty s^n \hat{U}(r) |n\rangle \langle n| \hat{U}^\dagger(r).
\end{eqnarray}
In general, the operator $\hat{\Gamma}_0$ can be interpreted as a density operator: It is positive semi-definite and $\text{tr}\hat{\Gamma}_0=1$. For the case at hand, the $\hat{\Gamma}_0$ of Eq. \eqref{eq:Gamma0Amp2} is a squeezed thermal state, with squeezing parameter $r=\ln \left(2\sigma^2+1\right)^{1/4}$, mean thermal photon number $\bar{n}=\frac{\sqrt{2\sigma^2+1}}{2} - \frac{1}{2}$, $s = \frac{\bar{n}}{\bar{n}+1}$, and $U(r)$ is the single-mode squeezing operator defined as,
\begin{eqnarray}
\nonumber \hat{U}(r)\hat{a}\hat{U}^\dagger(r) &\rightarrow& \begin{pmatrix}
	\hat{b} \\ \hat{b}^\dagger
	\end{pmatrix} = \begin{pmatrix}
	\cosh r & -\sinh r \\
	-\sinh r & \cosh r
	\end{pmatrix} \begin{pmatrix}
	\hat{a} \\ \hat{a}^\dagger
	\end{pmatrix},\\
\nonumber \hat{U}^\dagger(r)\hat{a}\hat{U}(r) &\rightarrow&	\begin{pmatrix}
	\hat{b} \\ \hat{b}^\dagger
	\end{pmatrix} = \begin{pmatrix}
	\cosh r & \sinh r \\
	\sinh r & \cosh r
	\end{pmatrix} \begin{pmatrix}
	\hat{a} \\ \hat{a}^\dagger
	\end{pmatrix}.
\end{eqnarray}

From Eq. \eqref{eq:Gamma2TP} we get, 
\begin{eqnarray}
\label{eq:HalfGaussianMoment}	\text{tr}\hat{\Gamma}_2 = \frac{2}{\sigma \sqrt{2\pi}} \int_0^{\infty} dq_\alpha q_\alpha^2 e^{-\frac{q_\alpha^2}{2\sigma^2}} = \sigma^2.
\end{eqnarray}

Since Eq. \eqref{eq:Gamma0Amp2} is the diagonal form of $\hat{\Gamma}_0$, Eqs. \eqref{eq:B} and \eqref{eq:Gamma0Amp2} give,
\begin{eqnarray}
\begin{aligned}
\hat{B}=& \frac{2}{1-s} \sum_{n,m=0}^{\infty} \frac{1}{s^n+s^m} \times
\\& \langle n |\hat{U}^\dagger (r) \hat{\Gamma}_1 \hat{U}(r) |m\rangle \hat{U}(r)|n\rangle \langle m|\hat{U}^\dagger (r) 
\end{aligned}
\end{eqnarray}
and
\begin{eqnarray}
\begin{aligned}
\text{tr}\left(\hat{B}\hat{\Gamma}_1\right) &= \frac{2}{1-s} \sum_{n,m=0}^{\infty} \frac{1}{s^n+s^m} \times 
\\& |\langle n |\hat{U}^\dagger (r) \hat{\Gamma}_1 \hat{U}(r) |m\rangle|^2.
\end{aligned}
\end{eqnarray}

Now let us evaluate the matrix element $\langle n |\hat{U}^\dagger (r) \hat{\Gamma}_1 \hat{U}(r) |m\rangle$ appearing in the previous expression. Equation \eqref{eq:GammakTP} for $k=1$ gives (for more details see Appendix \ref{app: FullExpr}),
\begin{eqnarray}
\nonumber \langle n |\hat{U}^\dagger(r) \hat{\Gamma}_1 \hat{U}(r) |m\rangle &=& \frac{1}{\sigma \sqrt{2\pi n! m!}} \frac{1}{\cosh{r}}\left(\frac{\tanh r}{2}\right)^{\frac{n+m}{2}}\\
\label{eq:amplitudeTP} &\times& \left(I_{nm}^{(+)} + I_{nm}^{(-)} \right),
\end{eqnarray}
where we used the projection of a single-mode displaced squeezed state on a Fock state \cite{Gong1990} and we denoted,
\begin{eqnarray}
\label{eq:integral}I_{nm}^{(\pm)} = \int^{+\infty}_0 dq_\alpha e^{-A q_\alpha^2} q_\alpha H_n(\pm x q_\alpha) H_m(\pm x q_\alpha),
\end{eqnarray}
where $H_n(\cdot)$ is the Hermite polynomial, $A=(1/2)(\sigma^{-2}+1-\tanh{r})>0$ and $x=(2\sinh{2r})^{-1/2}>0$ are known parameters depending only on $\sigma$ since $r=\ln{(2\sigma^2+1)^{1/4}}$. 

\section{Measurements}\label{SEC: Measurement technology}

\subsection{Photon number resolving}
For photon number resolving (PNR) detection, the MSE $\Tilde{\delta}_{\text{PNR}}$ can be computed as follows,
\begin{eqnarray}
\label{eq:MSE_PNR}
\Tilde{\delta}_{\text{PNR}}=\int_0^{\infty} dq_{\alpha} \sum_{k=0}^{n} P(q_{\alpha})P(k|q_{\alpha})\left( \Pi_k-q_{\alpha} \right)^2,
\end{eqnarray}
where $k$ is the detected photon number, $P(q_{\alpha})$ is the prior probability, $P(k|q_{\alpha})$ is the conditional probability, and $\Pi_k$ is the expected value,
\begin{eqnarray}
\nonumber \Pi_k&=&\int_0^{\infty} dq_{\alpha} P(q_{\alpha}|k) q_{\alpha}
\label{eq:MSE_pik1}\\&=& \frac{\int_0^{\infty} dq_{\alpha}  P(k|q_{\alpha})P(q_{\alpha}) q_{\alpha}}{\int_0^{\infty} dq_{\alpha} P(k|q_{\alpha})P(q_{\alpha})},
\end{eqnarray}
where we used Bayes' rule for $P(q_{\alpha}|k)$ and Born's rule to find,
\begin{eqnarray}
\nonumber P(k|q_{\alpha})&=&\langle k|\hat{\rho}|k\rangle\\
\nonumber &=&\langle k|\left(\frac{1}{2}|\alpha\rangle\langle \alpha |+\frac{1}{2}|-\alpha\rangle\langle -\alpha |\right) |k\rangle\\
\label{eq:BornsRule} &=&e^{-\frac{q^2_{\alpha}}{2}}\frac{(\frac{q_{\alpha}^2}{2})^{k}}{k!}.
\end{eqnarray}
From Eqs. \eqref{eq:MSE_PNR}, \eqref{eq:MSE_pik1}, and \eqref{eq:BornsRule} we get,
\begin{eqnarray}
\nonumber \tilde{\delta}_{\text{PNR}} &=& \sigma^2-\frac{2 \sigma^2}{\pi \sqrt{\sigma^2+1}}\\
\label{eq:MSE_PNR2}&\times& \left[\sigma \arcsin\left(\frac{\sigma}{\sqrt{\sigma^2+1}}\right)+1\right].   
\end{eqnarray}

\subsection{Homodyne detection}
\begin{figure}[t]
\centering
\includegraphics[width=0.47\textwidth]{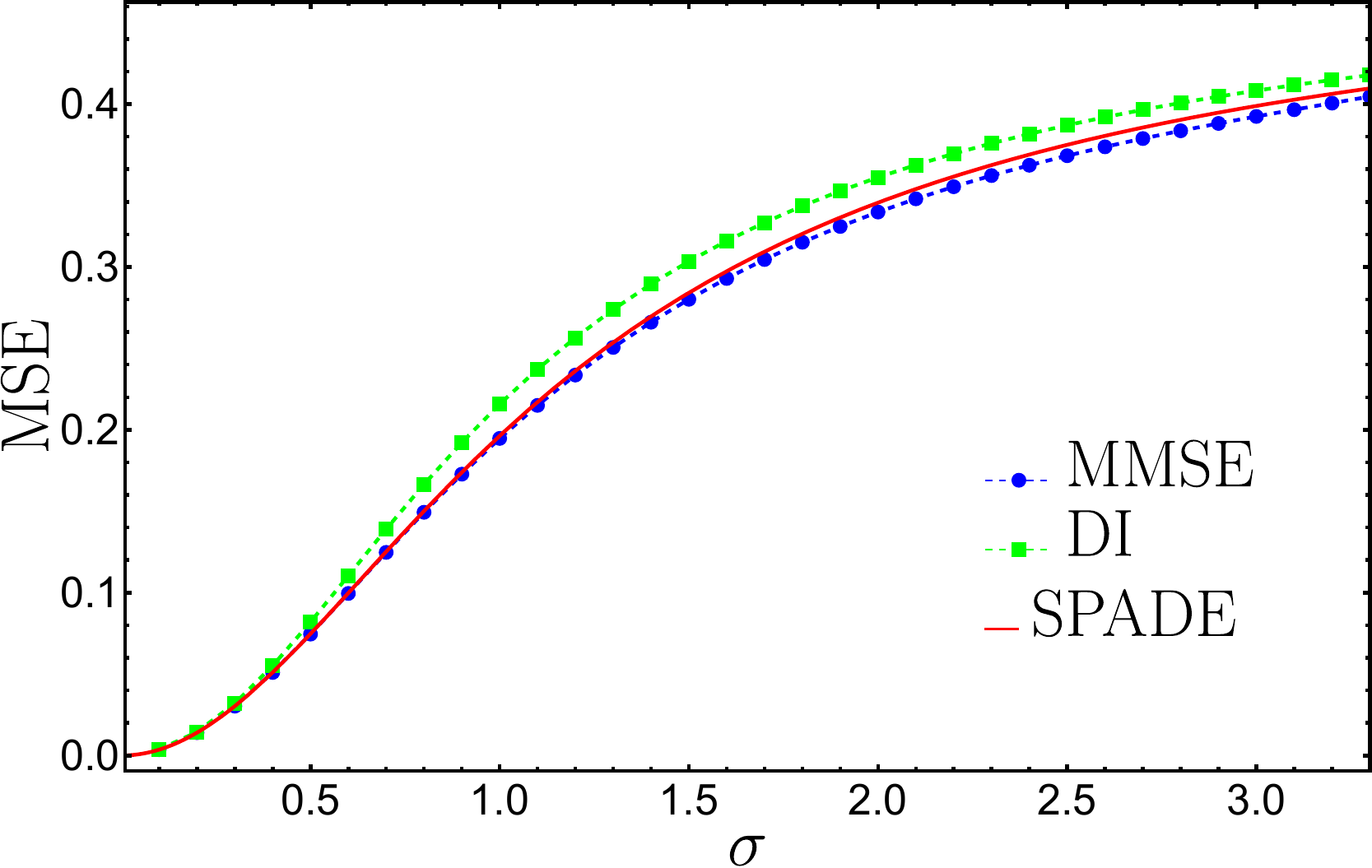}
\caption{Comparison of the MMSE, MSE for SPADE and MSE for DI using the half-Gaussian prior PDF. MSE for SPADE is closer to the MMSE, however it does not exhibit much advantage compared to the MSE for DI.}
\label{FIG: MSE compare}
\end{figure}
For homodyne detection, the MSE $\Tilde{\delta}_{\text{H}}$ can be computed as follows,
\begin{eqnarray}
\label{eq:MSE_HOM}
\tilde{\delta}_{\text{H}}=\int_0^{\infty} dq_{\alpha} \int_{-\infty}^{\infty} dq P(q_{\alpha})P(q|q_{\alpha})\left( \Pi_{q}-q_{\alpha} \right)^2,    
\end{eqnarray}
where the quadrature $q$ is the position and
\begin{eqnarray}
\nonumber \Pi_{q}&=&\int_{0}^{\infty}dq_{\alpha} P(q_{\alpha}|q) q_{\alpha}\\
&=&\frac{\int_0^{\infty} dq_{\alpha}  P(q|q_{\alpha})P(q_{\alpha}) q_{\alpha}}{\int_0^{\infty} dq_{\alpha} P(q|q_{\alpha})P(q_{\alpha})},
\end{eqnarray}
where we used Bayes' rule for $P(q_\alpha|q)$ and Born's rule to find,
\begin{eqnarray}
\nonumber P(q|q_{\alpha})&=&\langle q| \hat{\rho}(q_{\alpha}) |q \rangle\\
&=&\frac{e^{-(q-q_{\alpha})^2}+e^{-(q+q_{\alpha})^2}}{2\sqrt{\pi}}.
\end{eqnarray}
The expression for $\Tilde{\delta}_{\text{H}}$ cannot be found analytically. Therefore, we compute Eq. \eqref{eq:MSE_HOM} numerically. We compare the MMSE, PNR, and homodyne in Fig. \ref{FIG: MSE compare}, where, as we explain in Section \ref{sec:Spaces}, PNR and homodyne in the mathematical formalism we are utilizing, correspond to SPADE and direct imaging respectively in the imaging space.

\subsection{Imaging and hypothetical spaces}\label{sec:Spaces}
For the purpose of explanation and for this subsection only, we denote by I the imaging space described by the density operator $\hat{\rho}_I= w_1 |\psi_1\rangle \langle \psi_1|+w_2|\psi_2\rangle \langle \psi_2|$, and by H the hypothetical space described by the density operator $\hat{\rho}_H= w_1 |\alpha_1 \rangle \langle \alpha_1|+w_2 |\alpha_2 \rangle \langle \alpha_2|$. 

In the imaging space, the state of each point source can be expanded in the HG modes as $|\psi_i\rangle=\sum_{j=0}^{\infty}e^{-\frac{1}{2}|\frac{\chi}{2\sigma}|^2}\frac{(\frac{\chi}{2\sigma})^{j}}{\sqrt{j!}}|\phi_j\rangle$, where a single photon exists on infinite HG modes $|\phi_j\rangle$. In the hypothetical space, $|\alpha_i\rangle$ is just a coherent (single-mode) state which can be expressed on the (infinite) Fock basis $\{|j\rangle\}$. 
\begin{table*}
\caption{\label{Table 1} Relation of imaging basis and hypothetical spaces. See also \cite{Bisketzi_2019}.}
\begin{ruledtabular}
\begin{tabular}{lll}
 / & Imaging  & Hypothetical\\
\hline
State & $\hat{\rho}_I= w_1 |\psi_1\rangle \langle \psi_1|+w_2|\psi_2\rangle \langle \psi_2|$ & $\hat{\rho}_H= w_1 |\alpha_1 \rangle \langle \alpha_1|+w_2 |\alpha_2 \rangle \langle \alpha_2|$ \\
\hline
\multirow{2}{*}{Description} & $|\psi_i\rangle=\sum_{j=0}^{\infty}e^{-\frac{1}{2}|\frac{\chi}{2\sigma}|^2}\frac{(\frac{\chi}{2\sigma})^{j}}{\sqrt{j!}}|\phi_j\rangle$& $|\alpha_i\rangle=\sum_{j=0}^{\infty}e^{-\frac{1}{2}|\alpha_i|^2}\frac{\alpha_i^{j}}{\sqrt{j!}}|j\rangle$ \\ &1 photon with infinite modes &1 mode with infinite photons  \\
\hline
SPADE / PNR & $P_j=\langle \phi_j|\hat{\rho}_I|\phi_j\rangle$ & $P_j=\langle j|\hat{\rho}_H|j\rangle$ \\
\hline
DI / HOM &  \makecell[l]{$P(x)=w_1 \Psi_{\text{PSF}}^2(x-\chi_1)$ \\$\quad \quad \ +w_2 \Psi_{\text{PSF}}^2(x-\chi_2)$} & $P(x)=w_1|\langle x|\alpha_1\rangle|^2+w_2|\langle x|\alpha_2\rangle|^2$ \\ 
\end{tabular}
\end{ruledtabular}
\end{table*}

Thus, the photon number (Fock) basis in the hypothetical space corresponds to the HG basis in the imaging space. The PDF of PNR detection in the hypothetical space is $P_j=\langle j|\hat{\rho}_H|j\rangle$ for the $j$-th photon number eigenstate. This is equivalent to the probability of detecting the photon in the $j$-th HG mode, i.e., $P_j=\langle \phi_j|\hat{\rho}_I|\phi_j\rangle$, in the imaging space. 
A similar connection can be also found between homodyne detection and DI. When detecting quadrature $q$, the resulting PDF is $P(q)=\langle q|\hat{\rho}_H|q\rangle=w_1|\langle q|\alpha_1\rangle|^2+w_2|\langle q|\alpha_2\rangle|^2$. The equivalent form in the imaging space is DI, i.e., the measurement outcome is the position of arrival $q$ of the photon in the image plane, given by the PDF $P(q)=\langle q|\hat{\rho}_I|q\rangle=w_1 \Psi_{\text{PSF}}^2(q-\chi_1) +w_2 \Psi_{\text{PSF}}^2(q-\chi_2)$. The correspondence between imaging and hypothetical spaces is summarized in Table \ref{Table 1}.

\section{MMSE, MSE for DI, and MSE for SPADE: Displaced half-Gaussian PDF}\label{SEC: Displaced half Gaussian PDF}
\begin{figure*}[t]
\centering
\includegraphics[width=0.95\textwidth]{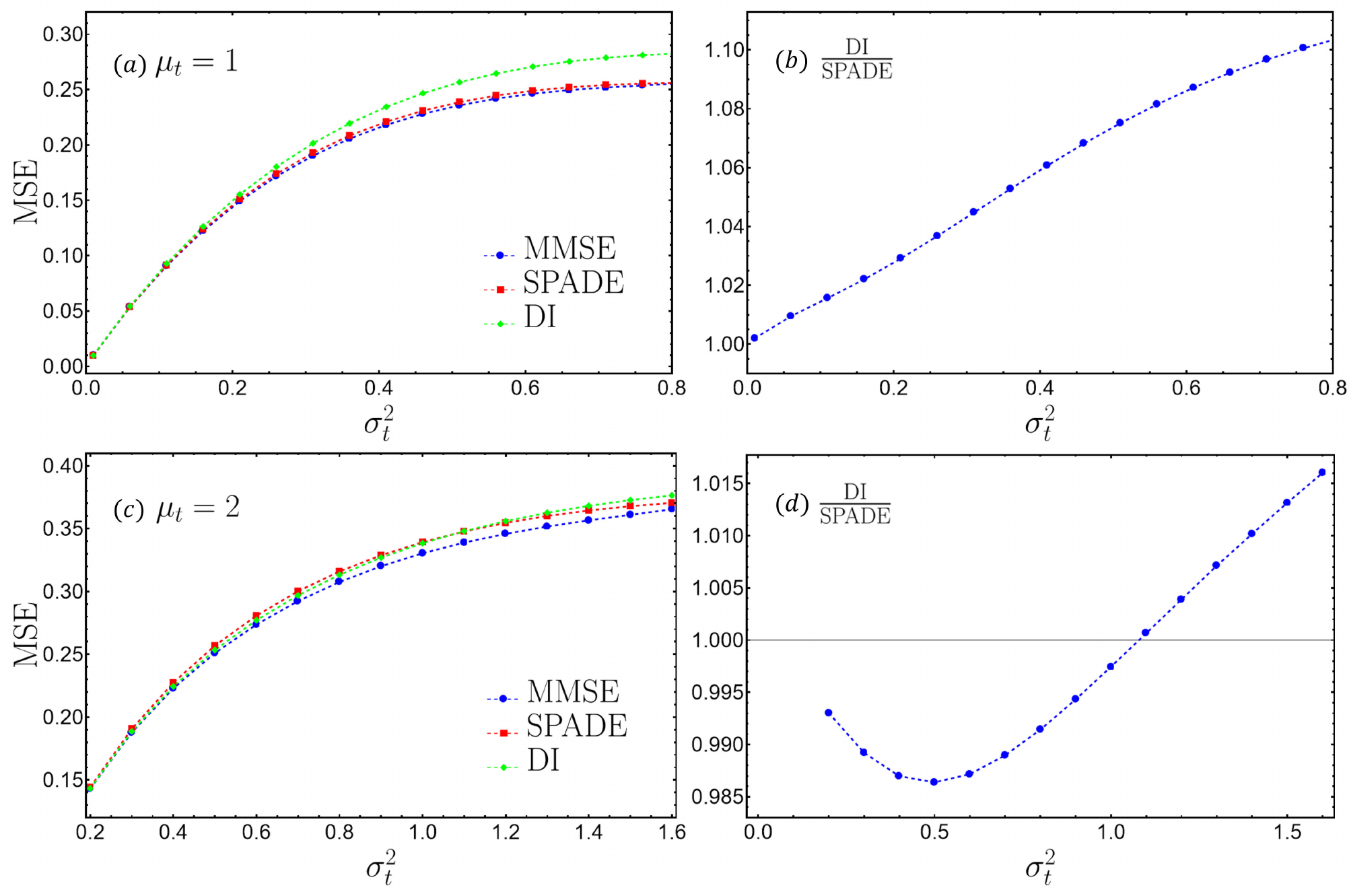}
\caption{The dependence of the MMSE, MSE for SPADE, and MSE for direct imaging (DI) on the variance $\sigma_t^2$ of the displaced half Gaussian prior PDF for constant mean $\mu_t$. True mean is, (a) and (b): $\mu_t=1$, (c) and (d): $\mu_t=2$. In (b) and (d) we show the ratio of MSE for DI and MSE for SPADE where when the ratio is below (above) $1$, DI is better (worse) than SPADE.}
\label{FIG: true mean}
\end{figure*}

\begin{figure*}[t]
\centering
\includegraphics[width=0.95\textwidth]{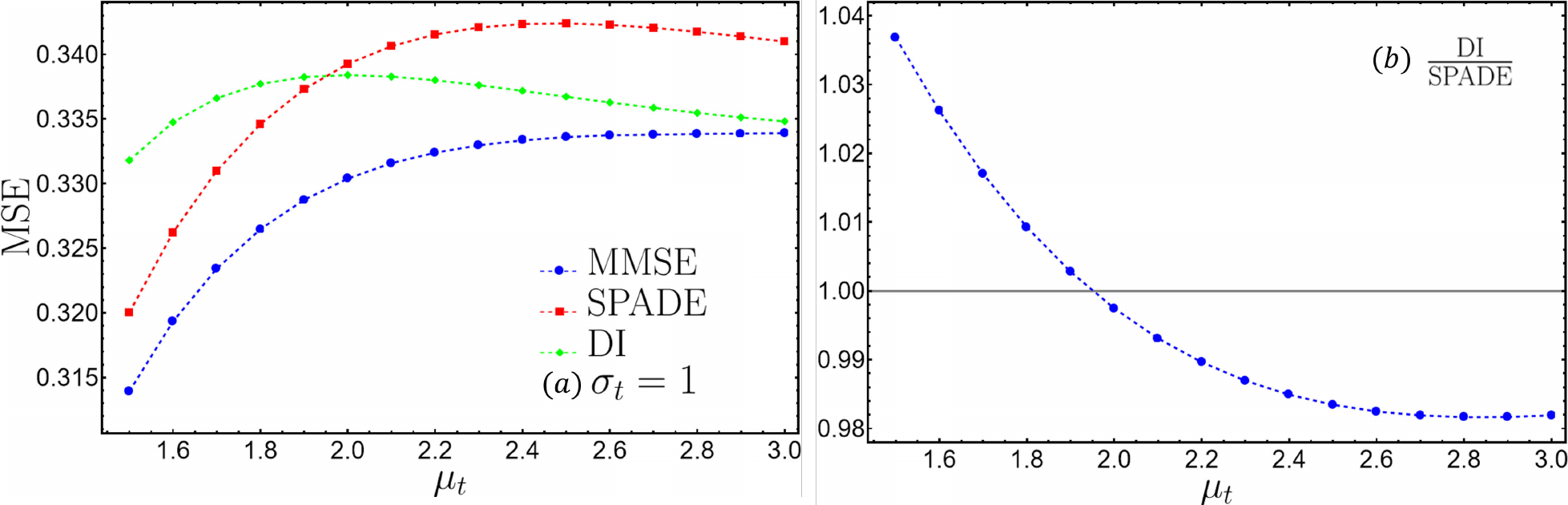}
\includegraphics[width=0.95\textwidth]{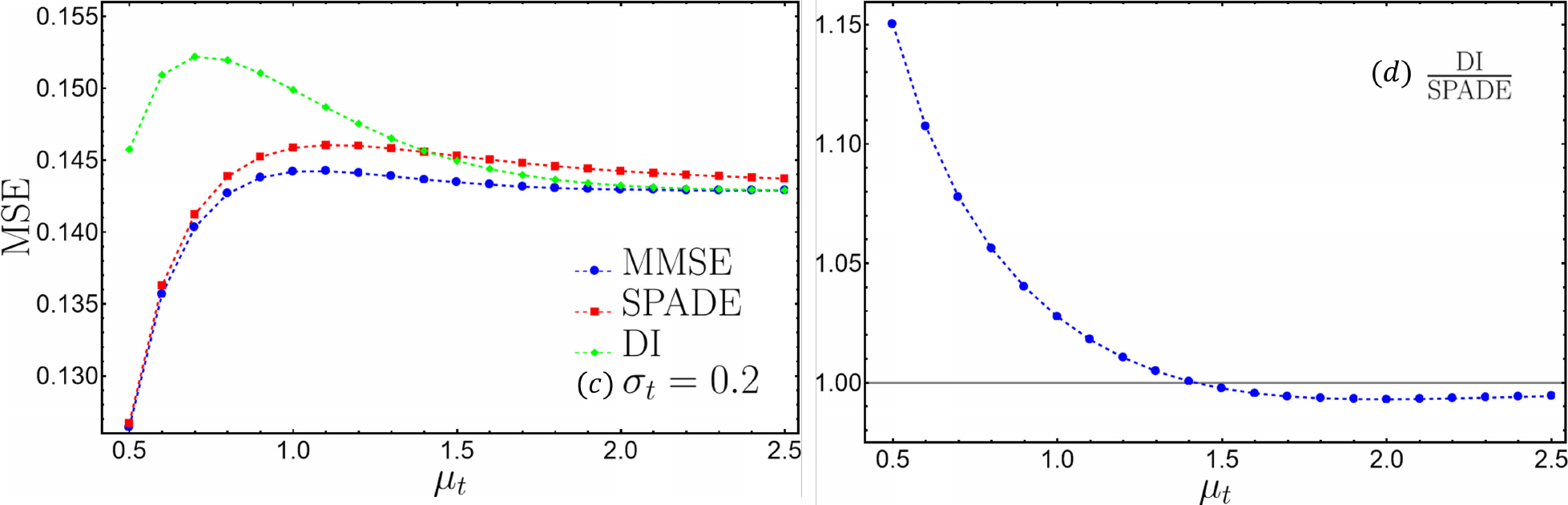}
\includegraphics[width=0.95\textwidth]{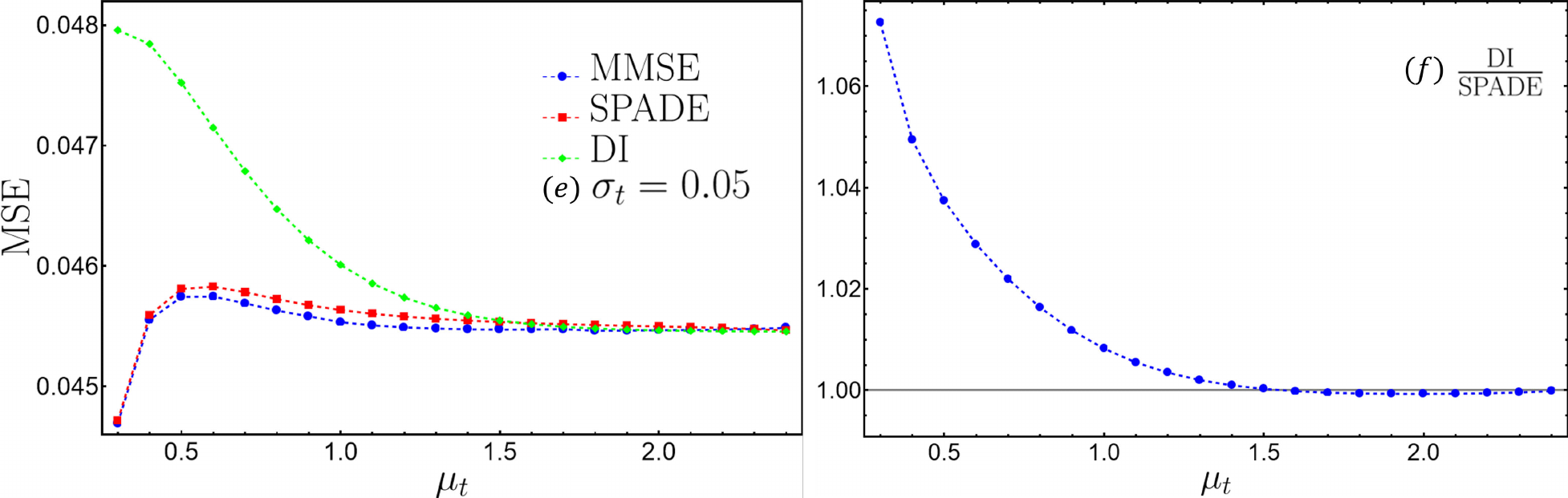}
\caption{The dependence of the MMSE, MSE for SPADE, and MSE for direct imaging (DI) on the mean $\mu_t$ for constant variance $\sigma_t^2$. Variance is, (a) and (b): $\sigma_t^2=1$,  (c) and (d): $\sigma_t^2=0.2$, (e) and (f): $\sigma_t^2=0.05$. In (b), (d), and (f) we show the ratio of MSE for DI and MSE for SPADE where when the ratio is below (above) $1$, DI is better (worse) than SPADE.}
\label{FIG: true variance}
\end{figure*}

The mean $\mu_t$ and variance $\sigma_t^2$ of the PDF of Eq. \eqref{eq:HalfGaussian} are,
$\mu_t=\sqrt{1/\pi}\sigma$, $\sigma_t^2=\left(1-2/\pi\right)\sigma^2$.
Therefore, for a given $\sigma$ the mean and variance of Eq. \eqref{eq:HalfGaussian} are not independent. For this reason we turn our attention to the displaced half Gaussian prior PDF,
\begin{eqnarray}
\label{eq:displaced half Gaussian PDF}
P(q_{\alpha})=\frac{1}{G}e^{-\frac{(q_{\alpha}-\mu)^2}{2\sigma^2}},
\end{eqnarray}
where the normalization $G$, the mean $\mu_t$ and variance $\sigma_t^2$ of Eq. \eqref{eq:displaced half Gaussian PDF} are,
\begin{eqnarray}
G&=&\sqrt{\frac{\pi}{2}}\sigma \left(1+\text{Erf}\left[\frac{\mu}{\sqrt{2}\sigma}\right]\right),\\
\label{eq:Mean displaced half Gaussian PDF}
\mu_t&=&\frac{\sqrt{\frac{2}{\pi }} \sigma  e^{-\frac{\mu ^2}{2 \sigma ^2}}}{\text{Erf}\left(\frac{\mu }{\sqrt{2} \sigma }\right)+1}+\mu,\\
\nonumber \sigma_t^2&=& \sigma^2-\frac{2 \sigma^2  e^{-\frac{\mu ^2}{\sigma ^2}}}{\pi  \left(\text{Erf}\left(\frac{\mu }{\sqrt{2} \sigma }\right)+1\right)^2}\\
\label{eq:Var displaced half Gaussian PDF} &+& \frac{\sqrt{\frac{2}{\pi }} \mu \sigma  e^{-\frac{\mu ^2}{2 \sigma ^2}}}{\text{Erfc}\left(\frac{\mu }{\sqrt{2} \sigma }\right)-2},
\end{eqnarray}
where $\text{Erf}(\cdot)$ is the error function and $\text{Erfc}(\cdot)$ is the complementary error function. 

We cannot obtain the analytical expressions of MMSE and MSE for DI and SPADE. Therefore, we use a numerical calculation.

The coherent state can be expanded in the Fock (number) basis. Using Eqs.~\eqref{eq:Gammak} and \eqref{eq:displaced half Gaussian PDF}, we can get $\hat{\Gamma}_k$ on the Fock basis,
\begin{widetext}
\begin{eqnarray}
\nonumber \hat{\Gamma}_k &=& \frac{1}{G}\int_0^{\infty} dq_\alpha q_{\alpha}^k e^{-\frac{(q_\alpha-\mu)^2}{2\sigma^2}} \left(\frac{1}{2}|\alpha\rangle\langle \alpha |+\frac{1}{2}|-\alpha\rangle\langle -\alpha |\right)
\\ \nonumber &=&\frac{1}{2G}\int_0^{\infty} dq_\alpha q_{\alpha}^k e^{-\frac{(q_\alpha-\mu)^2}{2\sigma^2}} \left( \sum_{n,m}e^{-|\alpha|^2}\frac{\alpha^{n+m}}{\sqrt{n! m!}}  |n\rangle\langle m |+ \sum_{n,m}e^{-|\alpha|^2}\frac{(-\alpha)^{n+m}}{\sqrt{n! m!}}  |n\rangle\langle m |  \right)
\\&=&\frac{1}{2G}\int_0^{\infty} dq_\alpha \sum_{n,m} q_{\alpha}^k e^{-|\frac{q_{\alpha}}{\sqrt{2}}|^2-\frac{(q_\alpha-\mu)^2}{2\sigma^2}}\frac{(\frac{q_{\alpha}}{\sqrt{2}})^{n+m}+(-\frac{q_{\alpha}}{\sqrt{2}})^{n+m}}{\sqrt{n! m!}}  |n\rangle \langle m| ,
\label{eq: Gammak in Fock basis}
\end{eqnarray}
\end{widetext}
where we choose the same setting $q_{\alpha}=\sqrt{2}\alpha$ as in Sec.~\ref{SEC: MMSE of two incoherent optical point sources}. We can easily simplify the equation and find that $\text{tr}\hat{\Gamma}_2=\sigma^2$. For $k=0,1$, we will resort to numerical evaluation by truncating the Fock expansion up to a cutoff where the MMSE converges. The cutoff depends on the parameters of the prior PDF, e.g., for Fig.~\ref{FIG: true mean} the maximum truncated number is $35$. 

In Fig.~\ref{FIG: true mean}, we fix $\mu_t$ and vary $\sigma_t^2$, while in Fig.~\ref{FIG: true variance} we fix $\sigma_t^2$ and vary $\mu_t$. This is possible by choosing the parameters $\mu$ and $\sigma$ accordingly. In Fig.~\ref{FIG: true mean} (a), or when the mean is even smaller $\mu_t <1$, the performance of SPADE is very close to optimal, that is, close to the MMSE. Therefore, in our setting, SPADE still has the advantage in the small separation regime, similarly to the Fisherian approach \cite{PhysRevX.6.031033}. 

In Fig.~\ref{FIG: true mean} (b), we plot the ratio of MSEs for DI and SPADE. It shows that for larger variance and small separation, the performance of SPADE is more pronounced. 

In Figs.~\ref{FIG: true mean} (c) and (d), $\mu_t=2$. In this range, SPADE has advantage when the variance is larger than $1.1$. Unlike the Fisherian approach, in our setting, we find that SPADE is not always optimal when the mean is large. If we keep increasing the variance, the MSE for SPADE is again smaller than the MSE for DI.

Increasing the variance will also increase the MMSE and MSEs. We find similar behavior in both the Bayesian and the Fisherian approaches: The SPADE measurement has advantage in the small separation, especially when the variance is small, SPADE can reach the lower bound, i.e., the MMSE. However, in the middle range, DI is better which is in contrast to the Fisherian approach. When the mean keeps increasing, all the measurements will get the same MSE which approaches the MMSE (see Fig.~\ref{FIG: true variance}).

\section{Conclusion}\label{SEC: Conclusion}

In this work we revisited the task of estimating the separation of two incoherent sources from the Bayesian perspective. We started by considering the half-Gaussian prior PDF. Then, in order to be able to analyze the task better we considered the displaced half-Gaussian PDF whose mean (variance) can be fixed while its variance (mean) varies.

Our results indicate similarities and differences with the Fisherian approach \cite{PhysRevX.6.031033}: In the Fisherian approach, SPADE can attain the QFI while the quantum Cram\'er-Rao bound is constant for any separation. Thus, SPADE can overcome the Rayleigh limit for small separation, while for large separation, SPADE and DI have the same performance. In our setting, i.e., for the prior PDF of Eq. \eqref{eq:displaced half Gaussian PDF}, the situation is more involved: Firstly, the MMSE is not constant as it naturally depends on the parameters of the prior PDF. Secondly, SPADE is superior to DI in the region of small mean, and the MSE for SPADE is close to the MMSE as the mean and variance become smaller. However, we note that SPADE is not the optimal measurement in general. Thirdly, in the middle range of $\mu_t$, we find that DI starts to outperform SPADE. Lastly, when the mean is large enough, the separation is also large. In this case, both SPADE and DI approach the MMSE, which gets us back to the classical case, i.e., when the Rayleigh limit is irrelevant as the two sources are far from each other.

Future research directions could include Bayesian adaptive techniques \cite{Zhou2024} for estimating the separation, simultaneous estimation \cite{Rubio2020} of centroid and separation, and focusing on active or passive Bayesian sensing pertaining to microscopy, and elaborating on the optimal measurement in the Bayesian approach.

\appendix

\section{Single-point source}\label{App: Single source}
In this appendix, we find the analytical expression for the case of a single-point source when estimating the real amplitude of a coherent state. We note that this derivation can be found in \cite[ Ch. VIII, Sec. 3 therein]{helstrom1976Book} and partially in \cite{Personick1971}. We did our own derivation which is slightly different than these found in the aforementioned references. We verified that when taking into account that $\hbar=1$, our result is identical to the one in \cite{Personick1971,helstrom1976Book} for the limiting case where the thermal photon number of the state used in \cite{Personick1971,helstrom1976Book} goes to zero. 

The prior PDF $P(q_\alpha)$ is a zero-mean Gaussian with known variance $\sigma$, $P(q_\alpha) = \frac{1}{\sigma \sqrt{2\pi}} e^{-\frac{q_\alpha^2}{2\sigma^2}}$. Letting $p_{\alpha}=0$, $\hat{\Gamma}_k$ is,
\begin{eqnarray}
\label{eq:GammakAmp1}
\begin{aligned}
\hat{\Gamma}_k =& \frac{1}{\sigma \sqrt{2\pi}} \int dq_\alpha q_\alpha^k e^{-\frac{q_\alpha^2}{2\sigma^2}} 
\\& \Big|\alpha=\frac{q_\alpha}{\sqrt{2}}\Big\rangle \Big\langle \alpha= \frac{q_\alpha}{\sqrt{2}}\Big|.    
\end{aligned}
\end{eqnarray}

$\hat{\Gamma}_0$ would be a thermal state if the prior distribution was $P(q_\alpha,p_\alpha)=(1/2 \pi \sigma^2) e^{-(q_\alpha^2+p_\alpha^2)/(2\sigma^2)}$, i.e., if we integrated over both real and imaginary parts of an $\alpha \in \mathbb{C}$. For our case we can rewrite Eq. \eqref{eq:GammakAmp1} as,

\begin{eqnarray}
\label{eq:GammakAmp2}
\begin{aligned}
\hat{\Gamma}_k =& \frac{1}{\sigma \sqrt{2\pi}} \int dq_\alpha dp_\alpha (\sqrt{2}\alpha)^k e^{-\frac{q_\alpha^2}{2\sigma^2}}
\\& \times \delta(p_\alpha)|\alpha\rangle \langle \alpha|, 
\end{aligned}
\end{eqnarray}
where $\alpha=\frac{1}{\sqrt{2}}(q_\alpha+i p_\alpha)$  and $\delta(p_\alpha)$ is the Dirac delta function, i.e., 
\begin{eqnarray}
	\delta(p_\alpha) = \lim_{\sigma \rightarrow 0} \frac{1}{\sigma\sqrt{2\pi}}e^{-\frac{p_\alpha^2}{2\sigma^2}}
\end{eqnarray}

From Eq. \eqref{eq:GammakAmp2} we observe that,
\begin{eqnarray}
\label{eq:trick}	\hat{\Gamma}_k = \sqrt{2}^k\hat{a}^k \hat{\Gamma}_0 = \sqrt{2}^k \hat{\Gamma}_0 \hat{a}^{\dagger k},
\end{eqnarray}
where the last equality is valid because $p_\alpha$ will be set to zero by the Dirac delta.
Let us use Eq.~\eqref{eq:GammakAmp2} to write $\Gamma_0$,
\begin{eqnarray}
\label{eq:Gamma0Amp1}	\hat{\Gamma}_0 = \frac{1}{\sigma \sqrt{2\pi}} \int dq_\alpha dp_\alpha e^{-\frac{q_\alpha^2}{2\sigma^2}} \delta(p_\alpha)|\alpha\rangle \langle \alpha|
\end{eqnarray}
which represents a state whose $P$-function is,
\begin{eqnarray}
\label{eq:PG}	P_G(q_\alpha,p_\alpha) = \frac{1}{\sigma \sqrt{2\pi}} e^{-\frac{q_\alpha^2}{2\sigma^2}} \delta(p_\alpha).
\end{eqnarray}
We can get the Wigner representation $W_G(q_\alpha,p_\alpha)$ which corresponds to the $P$ function of Eq.~\eqref{eq:PG} using,
\begin{eqnarray}
\begin{aligned}
W_G(q_\alpha,p_\alpha) &= \frac{1}{\pi} \int dq_\beta dp_\beta P_G(q_\beta,p_\beta) 
\\& \times \exp[-|q_\alpha-q_\beta+i p_\alpha-i p_\beta|^2]
\\&= \frac{1}{2\pi\sqrt{\det V}} \times
\\& \exp\left[-\frac{1}{2} (q_\alpha,p_\alpha)^T V^{-1} (q_\alpha,p_\alpha)\right],
\end{aligned}
\end{eqnarray}
where $V$ is the covariance matrix,
\begin{eqnarray}
	V=\begin{pmatrix}
	\sigma^2+\frac{1}{2} & 0 \\
	0 & \frac{1}{2}
	\end{pmatrix}.
\end{eqnarray}
We can bring $V$ into the symplectic diagonal form $V=S V_D S^T$ with,
\begin{eqnarray}
\label{eq:SympSq}
\begin{aligned}
S &=\begin{pmatrix}
	(2\sigma^2+1)^{-\frac{1}{4}} & 0\\
	0 & (2\sigma^2+1)^{\frac{1}{4}}
	\end{pmatrix}
 \\& \equiv  \begin{pmatrix}
	e^{-r}& 0\\
	0 & e^{r}
	\end{pmatrix},
\end{aligned}
\end{eqnarray}

\begin{eqnarray}
	V_D &=& \begin{pmatrix}
	\frac{\sqrt{2\sigma^2+1}}{2} & 0\\
	0 & \frac{\sqrt{2\sigma^2+1}}{2} 
	\end{pmatrix}.
\end{eqnarray}

The symplectic matrix $S$ represents squeezing with parameter 
\begin{eqnarray}
\label{eq:r} r=(1/4) \ln \left(2\sigma^2+1\right)\geq 0, 
\end{eqnarray}
therefore $\hat{\Gamma}_0$ is a squeezed state thermal state, squeezed by $r$ and mean thermal photon number,
\begin{eqnarray}
\label{eq:thermalphotons}	\bar{n}=\frac{\sqrt{2\sigma^2+1}}{2} - \frac{1}{2}.
\end{eqnarray}
Now, we can rewrite $\hat{\Gamma}_0$ as,
\begin{eqnarray}
\label{eq:Gamma0Amp2_app} \hat{\Gamma}_0= (1-s)\sum_{n=0}^\infty s^n \hat{U}(r) |n\rangle \langle n| \hat{U}^\dagger(r),
\end{eqnarray}
where,
\begin{eqnarray}
\label{eq:s}	s = \frac{\bar{n}}{\bar{n}+1},
\end{eqnarray}
and $\hat{U}(r)$ is a the single-mode squeezing unitary operator. Specifically, the action of $\hat{U}(r)$ in the Heisenberg picture is,
\begin{eqnarray}
\begin{aligned}
&\hat{U}(r)\hat{a}\hat{U}^\dagger (r)  \rightarrow
\\&
    \begin{pmatrix}
	\hat{b} \\ \hat{b}^\dagger
	\end{pmatrix} = \begin{pmatrix}
	\cosh r & -\sinh r \\
	-\sinh r & \cosh r
	\end{pmatrix} \begin{pmatrix}
	\hat{a} \\ \hat{a}^\dagger
	\end{pmatrix},
\end{aligned}
\end{eqnarray}
\begin{eqnarray}
\begin{aligned}
\label{eq:Hein} &\hat{U}^\dagger (r)\hat{a}\hat{U}(r) \rightarrow
\\&
    \begin{pmatrix}
	\hat{b} \\ \hat{b}^\dagger
	\end{pmatrix} = \begin{pmatrix}
	\cosh r & \sinh r \\
	\sinh r & \cosh r
	\end{pmatrix} \begin{pmatrix}
	\hat{a} \\ \hat{a}^\dagger
	\end{pmatrix},
\end{aligned}
\end{eqnarray}
where from Eq. \eqref{eq:SympSq},
\begin{eqnarray}
\label{eq:cosh}	\cosh r= \frac{(2\sigma^2+1)^{\frac{1}{4}}+(2\sigma^2+1)^{-\frac{1}{4}}}{2},\\
\label{eq:sinh}	\sinh r= \frac{(2\sigma^2+1)^{\frac{1}{4}}-(2\sigma^2+1)^{-\frac{1}{4}}}{2}.
\end{eqnarray}
Then let us calculate $\text{tr}\hat{\Gamma}_2$. Using Eq.~\eqref{eq:trick},~\eqref{eq:Gamma0Amp2_app} we get,
\begin{eqnarray}
\begin{aligned}
\text{tr}\hat{\Gamma}_2 &= 2\text{tr} (\hat{a}^2 \hat{\Gamma}_0) 
\\&= 2(1-s) \sum_{n=0}^\infty s^n \langle n| \hat{U}^\dagger(r) \hat{a}^2 \hat{U}(r) |n\rangle,    
\end{aligned}
\end{eqnarray}
by using Eq.~\eqref{eq:Hein},~\eqref{eq:thermalphotons},~\eqref{eq:s},~\eqref{eq:cosh}, and~\eqref{eq:sinh} we can get,
\begin{eqnarray}
\label{eq:trGamma2Amp}
	\text{tr}\hat{\Gamma}_2 = \frac{1+s}{1-s} \sinh 2r= \sigma^2,
\end{eqnarray}

Note that if we calculate the integral of Eq.~\eqref{eq:GammakAmp1} for $k=2$, we again find $\text{tr}\Gamma_2  = \sigma^2$. However, our approach allows us to go further with the calculation as we brought $\Gamma_0$ in the diagonal form on the basis $\{\hat{U}|n\rangle\}$. Indeed, we have,
\begin{eqnarray}
\label{eq:expGamma0Amp}
\begin{aligned}
&\exp[-\hat{\Gamma}_0 z] =
\\& \sum_{n=0}^\infty \exp\left[-(1-s)s^n z\right] \hat{U}(r)|n\rangle \langle n| \hat{U}^\dagger(r).    
\end{aligned}
\end{eqnarray}
Using Eqs. \eqref{eq:B}, \eqref{eq:trick}, and \eqref{eq:expGamma0Amp}, we get $\hat{B}$,
\begin{eqnarray}
\label{eq:BAmp}
\begin{aligned}
\hat{B} &= 2 \int_{0}^{\infty} d z e^{-\hat{\Gamma}_{0} z} \hat{\Gamma}_{1} e^{-\hat{\Gamma}_{0} z}
\\&= 2 \int_{0}^{\infty} d z  \sqrt{2}\sum_{n, m=0}^\infty \exp\left[-(1-s)(s^n+s^m)z\right] 
\\& \quad \times \hat{U}(r)|n\rangle \langle n|\hat{U}^\dagger(r) \hat{a} \hat{\Gamma}_0 \hat{U}(r) |m\rangle \langle m|\hat{U}^\dagger(r)
\\& =\frac{2\sqrt{2}}{1-s} \sum_{n, m=0}^\infty \frac{1}{s^n+s^m} \langle n |\hat{U}^\dagger(r) \hat{a} \hat{\Gamma}_0 \hat{U}(r) |m\rangle  
\\& \quad \times \hat{U}(r)|n\rangle \langle m|\hat{U}^\dagger(r).
\end{aligned}
\end{eqnarray}

We proceed by using Eq. \eqref{eq:trick} to find,
\begin{eqnarray}
\label{eq:trBGamma1Amp}
\begin{aligned}
\text{tr}(\hat{B}\hat{\Gamma}_1) &= \sqrt{2}\text{tr}(\hat{B}\hat{a}\hat{\Gamma}_0)
\\& = \frac{4}{1-s} \sum_{n, m=0}^\infty \frac{|\langle n |\hat{U}^\dagger(r) \hat{a} \hat{\Gamma}_0 \hat{U}(r) |m\rangle|^2}{s^n+s^m} 
\\& =  \frac{2\sigma^4}{1+2\sigma^2}.
\end{aligned}
\end{eqnarray}

We calculate the MMSE $\delta$ using Eq.~\eqref{eq:delta1},~\eqref{eq:trGamma2Amp}, and~\eqref{eq:trBGamma1Amp}
\begin{eqnarray}
\label{eq:MMSEAmp} \delta= \text{tr}(\hat{\Gamma}_2 - \hat{B}\hat{\Gamma}_1)= \frac{\sigma^2}{1+2 \sigma^2}.
\end{eqnarray}

We proceed to find the optimal measurement by finding the eigenvectors of $B$ given in Eq.~\eqref{eq:BAmp}. We define a new operator $\hat{\tilde{B}}$:
 \begin{eqnarray}
 \label{eq:BtildeAmp}	\hat{\tilde{B}} = \hat{U}^\dagger(r) \hat{B} \hat{U}(r).
 \end{eqnarray}
 We note that the eigenvectors of $\hat{B}$ are the eigenvectors of $\hat{\tilde{B}}$ transformed by $\hat{U}(r)$. Using Eqs.~\eqref{eq:r},~\eqref{eq:thermalphotons},~\eqref{eq:Gamma0Amp2_app},~\eqref{eq:s}, and~\eqref{eq:Hein}, we rewrite $\hat{\tilde{B}}$,
 \begin{eqnarray}
 \label{eq:BtildeAmp2}
 \begin{aligned}
\hat{\tilde{B}} &= \frac{\sqrt{2}\sigma^2}{(2\sigma^2+1)^{3/4}} \Big[\sum_{n=0}^\infty \sqrt{n+1} |n\rangle \langle n+1|
\\& \quad+\sum_{n=0}^\infty \sqrt{n}|n\rangle \langle n-1|\Big]
\\&= \frac{\sqrt{2}\sigma^2}{(2\sigma^2+1)^{3/4}} (\hat{a}+\hat{a}^\dagger)
\\&\Rightarrow \frac{2\sigma^2}{(2\sigma^2+1)^{3/4}}\hat{q},
 \end{aligned}
 \end{eqnarray}
 where $\hat{q}=\frac{1}{\sqrt{2}}(\hat{a}+\hat{a}^\dagger)$ is the position operator. Therefore, the eigenvectors of $\hat{\tilde{B}}$ are $\{|q\rangle\}$, where $\hat{q}|q\rangle=q|q\rangle$. From Eqs.~\eqref{eq:BtildeAmp} and~\eqref{eq:BtildeAmp2}, we conclude that the optimal receiver consists of a single mode squeezer $\hat{U}^\dagger(r)=\hat{U}(-r)$ acting on the output state followed by homodyne detection on position.
 
Here we make an observation: The MMSE is invariant under unitary transformations. This can be viewed by transforming $\hat{\Gamma}_k$  with $\hat{U}(r)$,
\begin{eqnarray}
\label{eq:GammakTilde} \hat{\tilde{\Gamma}}_{k} =\hat{U}^\dagger(r)\hat{\Gamma}_k \hat{U}(r),\ k=0,1,2.
\end{eqnarray}
Then, the new MMSE $\tilde{\delta}$ is,
\begin{eqnarray}
\nonumber \tilde{\delta} &=& \text{tr}\left(\hat{\tilde{\Gamma}}_{2}-\hat{\tilde{B}} \hat{\tilde{\Gamma}}_{1}\right)\\
\nonumber &=&\text{tr}\Big(\hat{U}^\dagger (r)\hat{\Gamma}_{2}\hat{U}(r)-
\hat{U}^\dagger (r) \hat{B} \hat{U}(r) \hat{U}^\dagger (r)\hat{\Gamma}_{1}\hat{U}(r)\Big)\\
\nonumber &=&\text{tr}\left(\hat{\Gamma}_{2}-\hat{B} \hat{\Gamma}_{1}\right)\\
&=& \delta.
\end{eqnarray}
The MMSE being invariant makes physical sense: The MMSE is not expected to change under a unitary operation (that does not depend on the unknown parameter) applied after the parameter has been imprinted on the state. Exploiting this type of invariance, one can consider that $\hat{U}(r)$ is applied just after the state picks up the unknown parameter. Then, said $\hat{U}(r)$ cancels out with $U^\dagger (r)$ derived as part of the optimal measurement (followed by homodyne detection on position). Therefore, the optimal measurement is homodyne detection on position, without any squeezing involved.

\section{Full expression of Eq.\eqref{eq:amplitudeTP}}\label{app: FullExpr}
\begin{widetext}
\begin{eqnarray}
\label{eq:amplitudeTP_Full}
\nonumber \langle n |\hat{U}^\dagger (r) \hat{\Gamma}_1 \hat{U}(r) |m\rangle &=& \frac{1}{\sigma \sqrt{2\pi}}
\int_{0}^{+\infty} dq_\alpha  e^{-\frac{q_\alpha^2}{2\sigma^2}} q_\alpha
\left[\langle n | \hat{U}^\dagger (r) |\alpha\rangle \langle \alpha|\hat{U}(r)|m\rangle
 +\langle n | \hat{U}^\dagger (r) |-\alpha\rangle \langle -\alpha|\hat{U}(r)|m\rangle \right]\\
\nonumber &=&  \frac{1}{\sigma \sqrt{2\pi n! m!}} \frac{1}{\cosh{r}}\left(\frac{\tanh r}{2}\right)^{\frac{n+m}{2}}
\int_{0}^{+\infty} dq_\alpha  e^{- A q_{\alpha}^{\CG{2}} } \CG{q_\alpha} \left[H_n(x q_\alpha)H_m(x q_\alpha) +H_n(-x q_\alpha)H_m(-x q_\alpha) \right]\\
&=& \frac{1}{\sigma \sqrt{2\pi n! m!}} \frac{1}{\cosh{r}}\left(\frac{\tanh r}{2}\right)^{\frac{n+m}{2}}
\left(I_{nm}^{(+)} + I_{nm}^{(-)} \right),
\end{eqnarray}
where $A=(1/2)(\sigma^{-2}+1-\tanh{r})$, $x=(2\sinh{2r})^{-1/2}$, and $r=\ln{(2\sigma^2+1)^{1/4}}$. Since $H_n(-x)=(-1)^n H_n(x)$, if $n+m=\text{odd}$, the integral is equal to $0$.

\end{widetext}

\bibliography{bibliogr.bib}

\end{document}